\documentclass[12pt,preprint]{aastex}

\shorttitle{Tidal Evolution and Transit Surveys} 
\shortauthors{Debes \& Jackson}
\begin{document}
\title{Too Little, Too Late: How the Tidal Evolution of Hot Jupiters affects Transit Surveys of Clusters}
\author{John H. Debes\altaffilmark{1,2}, Brian Jackson\altaffilmark{1,2}}

\altaffiltext{1}{Goddard Space Flight Center, Greenbelt, MD 20771}
\altaffiltext{2}{NASA Postdoctoral Program Fellow}

\begin{abstract}
The tidal evolution of hot Jupiters may change the efficiency of transit surveys of stellar clusters.  The orbital decay that hot Jupiters suffer may result in their destruction, leaving fewer transiting planets in older clusters.  We calculate the impact tidal evolution has for different assumed stellar populations, including that of 47~Tuc, a globular cluster that was the focus of an intense HST search for transits.  We find that in older clusters one expects to detect fewer transiting planets by a factor of two for surveys sensitive to Jupiter-like planets in orbits out to 0.5~AU, and up to a factor of 25 for surveys sensitive to Jupiter-like planets in orbits out to 0.08~AU.  Additionally, tidal evolution affects the distribution of transiting planets as a function of semi-major axis, producing larger orbital period gaps for transiting planets as the age of the cluster increases.  Tidal evolution can explain the lack of detected exoplanets in 47~Tuc without invoking other mechanisms.  Four open clusters residing within the {\em Kepler} fields of view have ages that span 0.4-8~Gyr--if {\em Kepler} can observe a significant number of planets in these clusters, it will provide key tests for our tidal evolution hypothesis.  Finally, our results suggest that observers wishing to discover transiting planets in clusters must have sufficient accuracy to detect lower mass planets, search larger numbers of cluster members, or have longer observation windows to be confident that a significant number of transits will occur for a population of stars.    
\end{abstract}

\keywords{planetary systems---globular clusters: individual (47~Tuc)---methods:numerical}

\section{Introduction}

Shortly after the discovery of the transiting exoplanet HD 209458b \citep{charbonneau}, a large scale survey for transiting exoplanets was conducted for the globular cluster 47~Tuc over 8 days using the WF/PC2 instrument on the Hubble Space Telescope (HST) \citep{gilliland00}.  The globular cluster 47~Tuc was predicted to be one of the better places to look for the signs of planet formation around pulsars due to its large stellar density and higher metallicity relative to other globular clusters \citep{sigurdsson92}.  The large core density made it feasible to search tens of thousands of stars for transits within the WF/PC2 camera field of view.  However, no planets were detected out of 34,000 stars, despite the expectation of roughly 17 transiting hot Jupiters.  The expectation was based on the assumption that 1\% of stars hosted hot Jupiters, such as is observed in the local Solar neighborhood.   More recently, transit surveys of the outer regions of 47 Tuc, where the stellar density is lower, have also been conducted \citep[][hereafter W05]{weldrake05}. W05 surveyed another 22,000 stars in the outer regions of 47 Tuc.  Similar to the HST survey, W05 found no evidence of transiting planets, even though W05 expected to detect seven planets under the same assumptions as the HST survey.

Explanations for the dearth of transits have invoked the paucity of metals in 47 Tuc \citep{fischer}, or the fact that the density of stars in the core of a globular cluster can disrupt most planetary systems  \citep{davies01,bonnell01,armitage00}.  

Amongst field stars, there is a well known correlation between the metallicity of the planet host star and the probability that it harbors a giant planet in an orbit with a period of $<$4 years \citep{fischer}.  If core accretion is the main mechanism by which planetary systems form, then a higher metallicity corresponds to a larger reservoir of planetary material available.  This larger reservoir may lead to a higher occurrence of planet formation.

47 Tuc, however, has a metallicity of [Fe/H]=-0.74 \citep{salaris98}, and the trend that \citet{fischer} presented only significantly extended out to [Fe/H]$\sim$-0.5.   If one extrapolates a power-law of the metallicity correlation  P$_{planet}=0.03[(N_{Fe}/N_{H})/(N_{Fe}/N_{H})_\odot]^{2.0}$\ to lower metallicities, then one would expect $\sim$0.1\% of stars to host planets in 47~Tuc.  More recent metal poor planet surveys  of stars with 0.0 $>$ [Fe/H] $>$ -1.0 suggest that the fraction of stars with planets trails to a constant occurrence of $\sim$1\% independent of metallicity, although this conclusion is not firm due to small number statistics for low metallicity stars \citep{sozzetti09}.   Hot Jupiters with periods $<$4.2 days, the maximum period detectable in the HST 47 Tuc survey, comprise about 17\% of detected exoplanets in the field, also apparently independent of metallicity.  Using these numbers, one would expect that the two 47 Tuc surveys, with a total of 56,000 stars observed, should have detected between 1 and 10 planets even with the cluster metallicity taken into account.  

Dynamical interactions in the globular cluster core, where stellar density is highest, can reduce the frequency of bound planetary systems through disruptive stellar encounters \citep{smith01,bonnell01}.  However, in the densest parts of 47~Tuc, planets that successfully form and migrate at an early time to orbits  with semi-major axes $<$0.05-0.1~AU should be safe from disruption for the lifetime of the globular cluster \citep{davies01,fregeau06}.  Dynamical interactions can remove some, but probably not enough planets to explain the null result in 47~Tuc, unless the density of systems severely inhibits planet formation of hot Jupiters in the first place \citep{armitage00,bonnell01,spurzem09}.  In particular, \citet{spurzem09} finds that a few percent of Jupiter-like planets with $a<5~AU$ will be removed through dynamical interactions with other stars over the current lifetime of 47~Tuc.

Another impact from the large density and size of a globular cluster is the intense EUV/FUV field that is present in the earliest times of the cluster from massive O stars.  \citet{armitage00} showed that if a cluster is larger than Orion, disks will be evaporated by the cluster UV field on timescales short relative to planet formation through core accretion; qualitatively consistent with the null result of the 47~Tuc survey.  Further and more detailed work has focused on the FUV field of smaller clusters, primarily because the distance at which a disk bearing star spends its time close to an O star sensitively affects disk destruction timescales and requires N-body calculations for a statistical study \citep{adams06}.  More detailed work would need to be done to quantify how much planet formation is suppressed in globular clusters.


However, \citet{weldrake05} searched a 52\arcmin $\times$ 52\arcmin\ field centered on the 47~Tuc core, where most of the stars observed would have experienced FUV fluxes comparable to the Orion cluster and still found a dearth of hot Jupiters.  It is likely that another explanation, rather than FUV erosion of disks, is required.

There is further evidence that metallicity, dynamical interactions, and FUV disk destruction are not sufficient to explain the deficit of planets seen in transit surveys of clusters.  Surveys of less dense open clusters have not had success in detecting transiting planets, despite younger, more metal rich clusters being searched \citep[e.g.][]{hartman09}.  Each survey typically has not observed a sufficient number of cluster members for a significant detection, but in aggregate the number of stars observed suggests there may be some difference in hot Jupiter survival or formation between field stars and stars in stellar clusters.

Previous studies have considered the destructive effects of tides
raised on hot Jupiters.  For example,  \citet{gu} considered tidal dissipation within close-in planets, and showed that, for a sufficiently large initial orbital eccentricity, the dissipation can inflate the planets' radii.  The inflation can cause the planets to lose their atmospheres through Roche lobe overflow. \citet{gu} suggested this process might account for the observed lack of very short period (less than 3 days) close-in planets around field stars, but this result also applies to the case of 47~Tuc.  The requirement of a large initial eccentricity is not probable if the majority of hot Jupiters form through migration to small semi-major axes in a gas disk \citep[cf.][]{lin96}.  Encounters between stars in dense clusters may excite the planets' eccentricities, perhaps leading to inflation and Roche lobe overflow. However, \citet{spurzem09} showed that the resulting eccentricities from such encounters are usually smaller than required.

Although tides raised on the surfaces of close-in planets likely play some role in removing planets, they are unimportant once the orbital eccentricity becomes small, and so may only act for a limited time. On the other hand, tides raised on a planet-hosting star will cause orbital decay long after eccentricity becomes small, eventually leading to the planet's destruction as it fills its Roche lobe or crashes into the star.   
 \citet{levrard09} showed that
the orbits of most transiting planets are unstable to orbital decay
resulting from the tide raised by the planets on their host stars.  Close in planets can raise significant tides on their host stars.  If the host star is rotating more slowly than the planetary companion, the tidal dissipation within the stellar photosphere casues the tidal bulge on the star to lag behind the planet.  This lag gives rise to a torque which removes orbital angular momentum from the planet and causes it to reduce its semi-major axis over Gyr timescales or less \citep{jackson09}. 
Thus, the explanation for the null result in 47 Tuc and the lack of open cluster
detections is that hot Jupiters have lifetimes shorter than a few Gyr,
and thus older star clusters have fewer hot Jupiters than a subset of
younger Þeld stars. 

\citet{jackson09} considered the observational effects of hot Jupiters
orbiting field stars which are tidally disrupted over time.  \citet{jackson09} suggested that this process might naturally account for the lack of very close-in planets around older stars.  Furthermore, planets around field stars have a positive correlation between the ages of planet-hosting stars and the orbital
distances of planets: hot Jupiters around older stars tend to be
more distant.  \citet{jackson09} showed that this correlation was
consistent with the predictions of orbital decay through the tide raised on the host star.
Since transiting planets tend to be closer to their host stars than non-transiting planets, transiting planets would be especially prone to tidal destruction \citep{jackson09}. Consequently, tidal destruction has important implications for transit surveys, especially of older stellar clusters, where tides have had ample time to act.  
		
In this paper we model the tidal evolution of hot Jupiters and determine the resulting impact on the frequency with which transiting planets are observed in clusters of stars.  We find that tidal evolution of hot Jupiters can explain the dearth of planets observed in 47~Tuc without additional explanations, and that tidal evolution can affect transit surveys of clusters in general.  

We discuss the calculations we make for our model in Section \ref{sec:model}.  In Section \ref{sec:general} we discuss our methodology for predicting the effects of tidal evolution on observed transiting planets in an idealized cluster of solar mass stars, while in Section \ref{sec:tuc} we focus on the 47~Tuc HST survey.  In Section \ref{sec:kepler} we investigate how tidal evolution could be tested by the {\em Kepler} Open Cluster Survey.  Finally we discuss our results in Section \ref{sec:disc}.

\section{Model}
\label{sec:model}
Orbital evolution due to tides and the resulting destruction of close-in planets can have an important influence on the observed distribution of close-in planets by clearing out many of the planets closest to their host stars. In order to evaluate this influence quantitatively, we model the orbital evolution of a large population of close-in planets over time. Then we consider how many of those planets could be observed to transit their host stars by a survey with given detection thresholds. We compare the number of planets that we would expect to be detected by such a survey if tidal destruction of close-in planets is included to the number expected if tides had no influence. In this way, we quantitatively predict the expected number of transiting planets in a stellar cluster. We also tailor our calculations to specific stellar clusters, such as 47~Tuc, by considering distributions of stellar masses, radii and ages.

In developing a model population of planetary systems, we first select stellar masses $M_*$, radii $R_*$, and magnitudes $V_*$ for a population of planet-hosting stars (generally $10^6$ stars). In some of our model calculations, we assume all host stars in the model population have the same $M_*$, $R_*$, and $V_*$, while in other calculations, we choose a distribution of stellar parameters, based on estimates of stellar properties for real clusters. The ways in which we chose stellar masses, radii and magnitudes for each calculation are described in the subsequent sections. These stellar parameters together help determine the rate of orbital evolution and the observability of a planetary transit, as discussed below. For all stars in a model cluster, we fixed the age of all the planetary systems at a given value. We also fix all planetary masses $M_p$ at Jupiter's mass, and the radii $R_p$ at 1.2 Jupiter radii, the average radius for all transiting exoplanets. The value of $M_p$ helps determine a planet's orbital evolution rate, while the value of $R_p$ helps determine a planet's transit depth and thus its detectability. 

Next we select an initial semi-major axis value $a_{init}$ and an orbital inclination $i$ (the angle between the orbit normal and the observer's line-of-sight) for each planet in our model population. The $a_{init}$-value is selected at random, with a uniform probability to lie between $R_*+R_p$ (usually about 0.005 AU) and 0.5 AU.  The probability that the transit of a planet in a circular orbit with a semi-major axis larger than 0.5 AU will be visible from Earth is less than about 1\%, and orbital decay from tides has little influence on planets so far from their host stars. Thus we do not include planets with $a_{init} > 0.5$ AU.  Recent studies suggest the distribution of exoplanet semi-major axes may not be uniform \citep{cumming08}, but experimentation with non-uniform distributions show that our results are insensitive to our choice of $a_{init}$ distribution, as long as the distribution is not pathological (\emph{e.g.}, all $a_{init}$ = 0.01 AU). The $i$-value for each planet is selected at random, with a uniform probability to lie between 0 and 90$^{\circ}$.  Planetary systems with $i$-values not favorable for transit viewing are still followed throughout the full tidal evolution in order to record planets that later migrate into observable geometries. 

We next evaluate how many of our model planets we would expect a transit survey to detect.  
The following inequality must be satisfied for a planet to transit its host star: 

\begin{equation}
a \sin{i} \le R_* + R_{\rm p}
\label{eqn:Ptrans}
\end{equation}

where $a$ is the orbital semi-major axis.  

For some of our model calculations, we next determine whether the transits observable from Earth would produce a signal large enough to be detected.  
We use the same criteria as was used in the HST 47~Tuc survey, namely the detection of at least two transits with a total S/N $>$6.3.  This selection is dependent on the total number of stars being searched to ensure $\leq$1 false transit detection in the entire sample.  

The significance of a detection for a survey with near continuous coverage and white noise is simply dependent on the depth of the transit $d_{\rm tran}$ and the length of the transit $\tau_{\rm tran}$ relative to the S/N of the target's photometry and the exposure time of the observations $\tau_{\rm exp}$.  For non-grazing transits, the depth of the transit is given by the ratio of the planet's disk area to the stellar disk area ($R_{\rm p}/R_\star)^2$.

The length of a transit assuming a given inclination is given by \citep{sackett99}:

\begin{equation}
\tau_{\rm tran} = \frac{P_{\rm orb}}{\pi}\left[\left(\frac{R_\star+R_{\rm pl}}{a}\right)^2-\cos{i}^2\right]^{\frac{1}{2}}
\end{equation}

where $P_{\rm {orb}}$ is the orbital period.   The significance of a transit detection would then be the depth of the transit relative to the photometric uncertainties multiplied by the square root of the number of transit samples:

\begin{equation}
\sigma_{\rm {tran}} = \frac{d_{\rm {tran}}}{\sigma_{\rm {phot}}}\left(N_{\rm {tran}}\frac{\tau_{\rm {tran}}}{\tau_{\rm {exp}}}\right)^{1/2}
\end{equation}

where N$_{\rm {tran}}$ is roughly given by the ratio of the observing window to the orbital period of the planet.  The above equations assume that the transit is not grazing, when only part of the planetary disk transits the star.  For the small subset of model grazing transits, we integrated the transit depth over the entire length of the transit to derive an average transit depth for our calculations.

To get a sense of the typical signifiance of a transit, we assume a 1~M$_{\rm {J}}$ planet with radius 1.2~R$_{\rm {J}}$ in a 3.4 day orbit and $i=90^\circ$ around a 1 M$_\odot$ and 1 R$_\odot$ star.  If the epoch to epoch S/N=200, its transits will be detected with a $\sigma_{\rm {tran}}$=15.4 with a 17 day observing window and exposures taken every 16 minutes.

The continuous sampling approach with white noise is sufficient for exploring space based surveys like that performed for 47~Tuc and what is currently being performed by {\em Kepler}, but ground-based surveys need to account for more complicated noise considerations and uneven sampling timescales, which is beyond the scope of this paper \citep[cf.][]{pont06}.

We next determine the orbital evolution of the planets from tides raised on the host stars. Using the age we've assumed for the model population, we determine each planet's $a$-value at that age, according to \citep{jackson09}:

\begin{equation}
\label{eq:tidal}
a(t) = \left[a_{init}^{13/2} - \frac{117}{4}\left(\frac{G}{M_*}\right)^{1/2}\frac{M_{\rm p} R_{\rm{ZAMS}}^5}{Q_*^{\prime}}t\right]^{2/13}
\label{eqn:aevol}
\end{equation}

where $t$ is the desired age, $G$ is the gravitational constant, $R_{\rm {ZAMS}}$ is the zero age main sequence stellar radius (which can be different from R$_\star$ if the cluster is old enough), and $Q_*^{\prime}$ is the modified stellar tidal dissipation parameter \citep[e.g.][]{ogilvie07}.  We assume that orbital migration from interactions with a protoplanetary disk or other mechanisms no longer acts on the planets.  

Equation \ref{eq:tidal} also assumes that perturbations from additional bodies are negligible and that orbits are circular and the orbital axis is aligned with the stellar rotation axis.  It also assumes that $R_{\rm ZAMS}$ remains constant over the timescale of the calculation.

Since we assume that the stellar radius is constant over the calculation, it is important to use $R_{\rm {ZAMS}}$ rather than $R_\star$ so that one does not overestimate the impact of tidal evolution for stars that have recently left the main sequence, i.e. when their radii are 2-3 times larger than R$_{\rm ZAMS}$.  This assumption breaks down for giant stars whose current radius is large enough that the amount of tidal evolution is larger on the giant branch than during the entire main sequence lifetime of the star \citep[e.g.][]{villaver09}.  Since most transit surveys focus on main sequence stars and $R_\star$ does not significantly change over the main sequence lifetime, the time evolution of $R_\star$ does not impact our tidal evolution calculations more than other uncertainties.

For all stars in a given population, we consider a fixed value of $Q_*^{\prime} = 10^6$ \citep{jackson08a}. Some recent studies suggest the $Q_\star^{\prime}$-value appropriate for stars hosting close-in planets may actually lie in the range $10^8$ to $10^9$ \citep{ogilvie07,penev09,hebb09}, while orbital circularization of binary stars suggest $Q_\star^{\prime} \sim 10^5$ \citep{mathieu94}.    We discuss these uncertainties further in Sections \ref{sec:general} and \ref{sec:disc}.

Once we've calculated the $a$-value for each planet at the desired age, we apply the same transit probability and S/N considerations discussed above to determine whether the planet would be observed in the new orbit.  We remove planets from the sample for which $a(t) < R_* + R_p$, assuming that those planets have either become engulfed within the stellar photosphere or suffered significant Roche lobe overflow such that the planet has either mostly or entirely lost its atmosphere, rendering too small to observe.  

Evidence for the loss of a planetary atmosphere due to Roche lobe overflow can be found in transiting planets in the field.  Recent results for WASP-12b, a highly irradiated hot Jupiter with a semi-major axis of 0.023~AU, show a significant metal rich exosphere that extends beyond the Roche radius of the planet \citep{fosatti10}.  These observations confirm predictions from \citet{li10} where they calculate a mass loss rate for the planet that suggests it will be tidally stripped in $\sim$10~Myr.  If more systems like WASP-12b are found, it would suggest a primary pathway for the tidal destruction of the planets we consider in our study.

\section{Impact of tidal evolution on a Cluster Transit Survey of Sun-like stars}
\label{sec:general}
We first investigate the evolution of a population of planets with 1.2~$R_J$, 1 M$_J$ in orbit around a cluster of 1~R$_\odot$, 1~$M_\odot$ stars.  For this calculation, we considered clusters with ages ranging from 1~Myr to 10~Gyr, spanning the ages of star formation regions to globular clusters like 47~Tuc.  We assumed two separate observing conditions:  the first is an idealized survey with an observing window of infinite length and no S/N constraints, the second with constraints similar to the 47~Tuc HST survey--an observing window of 8.4 days, S/N=200 for V=18.4 stars, and a $\sigma_{\rm {tran}}\ge$6.3 requirement for detection.

\begin{figure}
\plotone{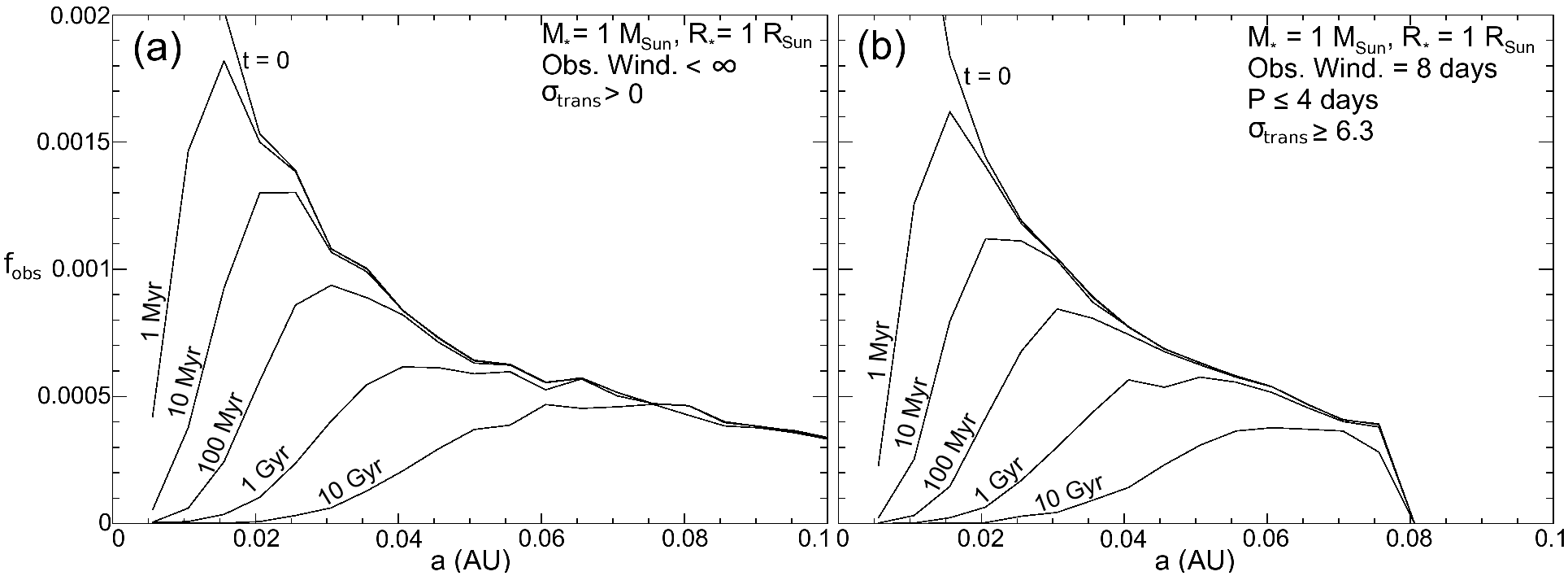}
\caption{\label{fig:f1}  This figure shows the evolution of the semi-major axis distribution of surviving, observable hot Jupiters around a cluster of 1 M$_\odot$, 1 R$_\odot$ stars as a function of cluster age for two sets of observing constraints: (left)$\sigma_{\rm {tran}}>$0 and an infinite observing window, and (right) observing constraints the same as the HST 47~Tuc survey: an 8.4 day observing window and $\sigma_{\rm {tran}}>$6.3, assuming a S/N=200 for a star with V=18.4.  The distribution can be compared to no tidal evolution with the t=0 line.  tidal evolution removes the planets with the highest probability to transit.}
\end{figure}

\begin{figure}
\plotone{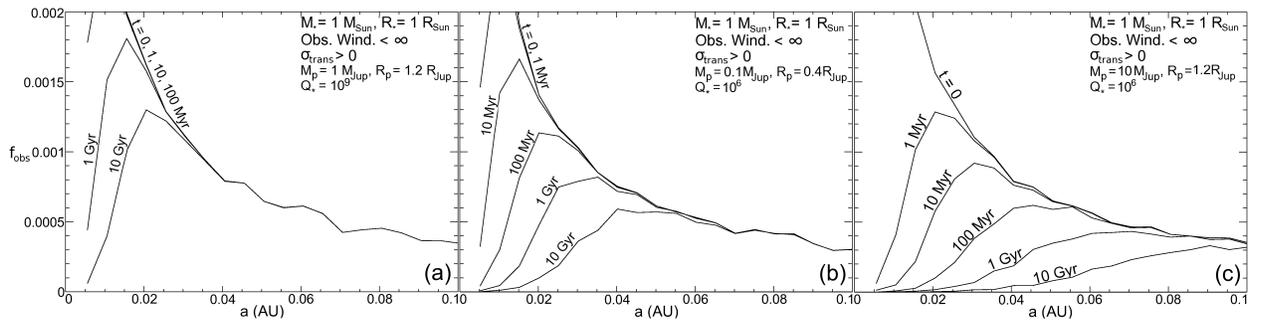}
\caption{\label{fig:f2} Same as for Figure \ref{fig:f1} for an idealized cluster survey but for different assumed parameters.  Panel a) shows the resulting semi-major axis distributions for a $Q_\star^\prime$=10$^9$, Panel b) shows the resulting distributions for target planets with M$_p$=0.1~M$_J$, and Panel c) shows the resulting distributions for target planets with M$_p$=10~M$_J$.}
\end{figure}

\begin{figure}
\plotone{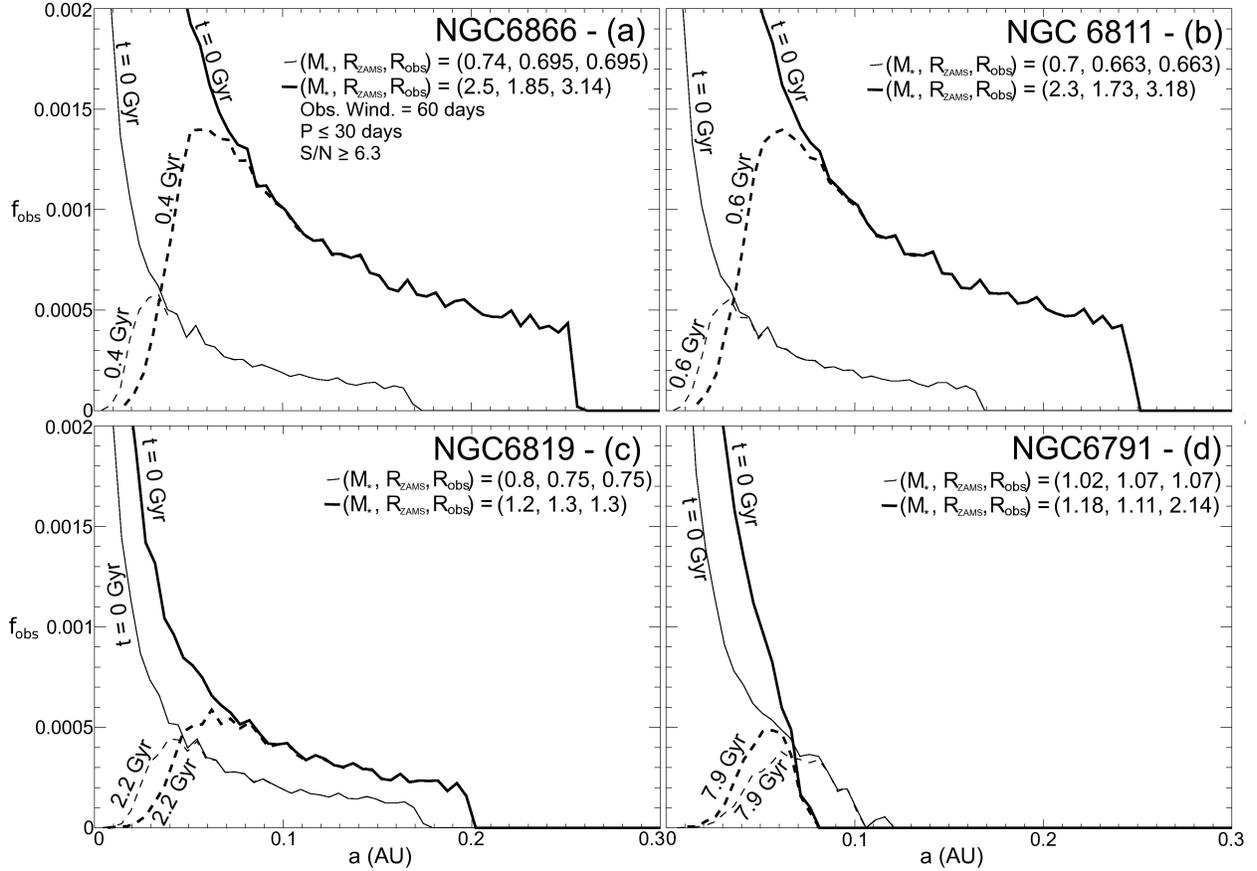}
\caption{\label{fig:f6} Expected semi-major axis distributions of observable hot Jupiters for each of the clusters within the {\em Kepler} field of view.  The solid lines are the initial orbital distributions, while the dashed lines are the distributions after tidal evolution for the age of the cluster. Bold lines are for the brightest main sequence cluster members, while thin dashed lines are for the dimmest main sequence cluster members accessible with {\em Kepler}.}
\end{figure}

The resulting distributions of detected planets as a function of semi-major axis are shown in Figure \ref{fig:f1}.  Panel (a) shows the idealized, S/N$>$0 distribution of planet semi-major axes as a function of cluster age.  The vertical axis gives the fraction of detected systems (f$_{obs}$ out of the 10$^6$ model planetary systems (also shown in Figures \ref{fig:f2},\ref{fig:f3}, and \ref{fig:f6}).  The horizontal bins are 0.005~AU wide.  To determine the number of predicted transiting planets in any bin for a given cluster, one must multiply $f_{obs}$ by $N_{\rm stars}$, where $N_{\rm stars}$ is the number of stars surveyed in a cluster expected to have planets.  The t=0 curves in both panels show the observed fraction of planets that transit in the absence of tidal evolution.  The number of planets observed decreases $\propto a^{-1}$, consistent with the expected geometry of a transit.  We denote this fraction f$_{obs,0}$.  For t$>$0, tidal interactions are strongest for shorter period planets, resulting in their prompt removal,.  A precipitous drop can be seen in transiting planets with semi-major axes $<$0.05~AU.  In panel (b), we see the resulting distribution for a survey with the same S/N threshold for detection and observing window as those of the HST 47~Tuc survey, which is most sensitive to planets with small semi-major axes.  In this case, the length of the observing window limits the periods of observable planets to $\sim$4 d, corrsponding to a semi-major axis of 0.08~AU.

Not all transiting planets will have a mass of 1~M$_J$, a radius of 1.2~R$_J$.  Similarly,  Q$_\star^\prime$ may not equal 10$^6$.  We investigated the impact of changing our assumptions by re-running the calculation in panel (a) of Figure \ref{fig:f1} with different $Q_\star^\prime$ and $M_{p}$.  For the first test we chose $Q_\star^\prime$=10$^9$, which lies at the extreme end of what is predicted by other studies \citep{ogilvie07,penev09,hebb09}.  The results of this simulation are given in panel (a) of Figure \ref{fig:f2}.  The main effect is a reduction in the efficiency with which tidal evolution removes hot Jupiters in close orbits, where as many as a factor of three more planets survive in orbits with semi-major axes of $\sim$0.02~AU.  If 10$^9$ is the correct value for Q$_\star^\prime$ then tidal evolution will be a less important consideration for transit surveys of clusters.

Our second and third tests varied the mass of the target planet, with one test using 0.1~M$_{J}$ and the second test using 10~M$_{J}$.  The resulting semi-major axis distributions from these simulations are shown in panel (b) of Figure \ref{fig:f2} for 0.1 M$_J$ and panel (c) for 10M$_J$.  The less massive planets raise smaller tides on their host stars and thus do not migrate in as quickly, but the difference in $f_{obs}$ is not much greater than a factor of two for planets in close orbits.  Larger planets, such as those in the second test, spiral in much more quickly, but again this modifies $f_{obs}$ by less than a factor of two for the closer orbits.

\section{Impact of tidal evolution on the 47 Tuc Survey}
\label{sec:tuc}
In the HST 47 Tuc Survey \citep{gilliland00}, the masses and radii of the observed stars spanned a broader range of values than considered in Section \ref{sec:general}.  Therefore, the impacts of tidal evolution may be different from our general examples above.  In the HST survey, the globular cluster was observed $\sim$6 times in the F555W and F814W filters for a total of twelve independent photometric points per $\sim$96 minute HST orbit.  The cluster was observed over 8.3 days, giving largely complete coverage of $\sim$40,000 stars, of which $\sim$34,000 were used to look for transiting planets \citep{gilliland00}.  

In order to reproduce the expected effects of tidal evolution on hot Jupiters on this survey, we determined the distributions of radii and masses of the stars in the HST sample.  To do so, we took the measured V magnitudes of the target stars (kindly provided by R. Gilliland) and converted them into masses and radii, interpolating over the values reported in \citet{gilliland00} and derived from isochrones of metal poor stellar models \citep{bergbusch92}.  

Since the higher mass stars targeted in 47~Tuc recently turned off the main sequence, we must infer the primordial $R_\star$ to accurately calculate the magnitude of tidal evolution on a particular orbit.  To determine $R_{\rm {ZAMS}}$ of these stars, we extrapolated the stellar radii for stars with masses $>$0.8 M$_\odot$ in 47~Tuc, assuming R$_\star \sim R_\star(0.8 M_\odot)*(M_\star/0.8 M_\odot)^{0.993}$ from the empirically determined masses and radii of eclipsing binaries \citep{gorda98}.

Just as in previous section, we ran Monte Carlo models to determine the semi-major axis distribution of observable transiting planets for 47~Tuc.  47~Tuc's most recent age estimate places it at 11.25~Gyr \citep{thompson09}.  Figure \ref{fig:f2} shows the distribution of planet semi-major axis as a function of time for 47~Tuc with our simulations.  Qualitatively, the idealized cluster described in Section \ref{sec:general}, especially under the same observing conditions, looks the same as for the 47~Tuc survey.  However, the dearth of hot Jupiters in the 1~M$_\odot$ cluster extends out to $\sim$0.055~AU, compared to 0.05~AU for 47~Tuc at 10~Gyr.  This extended inner hole is primarily because the stars in 47~Tuc are lower mass and thus have smaller radii than the Sun.  Less planets are removed in 47~Tuc from further radii than the larger Sun-like cluster at the same age.

\begin{figure}
\plotone{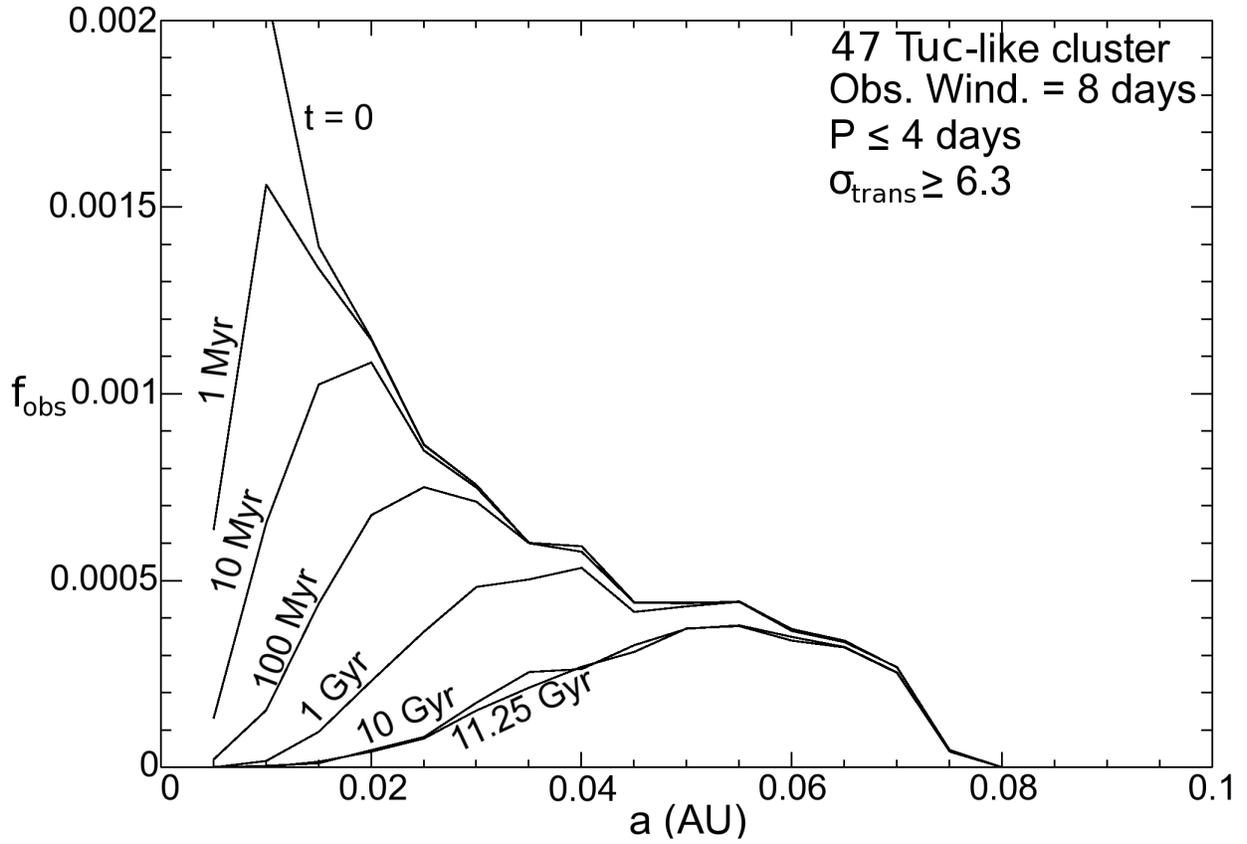}
\caption{\label{fig:f3} Same as for Figure \ref{fig:f1} but with a distribution of stellar masses and radii estimated to mimic that of the 47~Tuc cluster over cluster ages ranging from 1~Myr to 11.25~Gyr, the estimated age of 47~Tuc.  Removal of hot Jupiters through tidal evolution is sufficient to explain the null result of the 47~Tuc HST survey without needing any other mechanisms.}
\end{figure}

If we integrate over all observed semi-major axes in Figure \ref{fig:f3}, we can determine the overall fractional decrease in detected planets relative to a population where tides are not accounted for.  This fractional decrease, which we represent as $f_{obs}/f_{obs,0}$ has the advantage that it can be inserted into general expected planet yields of transit surveys in order to generate a more accurate estimate for a particular transit survey.   

Figure \ref{fig:f3} shows the overall percentage of planets that survive tidal evolution for the three different observing conditions: the idealized S/N$>$0 case for a sun-like cluster of stars (appropriate for surveys where the photometric accuracy of a transit survey is $\gg$d$_{\rm tran}$), the Sun-like cluster with an observing strategy similar to the HST 47~Tuc survey, and the 47~Tuc survey.  At 11.25~Gyr, tides remove all but 4\% of the original hot Jupiter population.  

If we combine the expectations of W05 and the HST 47~Tuc survey (17 from the HST survey, 7 from W05), tidal interactions can again account for the null result of the combined surveys without the need to invoke other explanations.  In reality, metallicity and the stellar density of the globular cluster must also impact the expected number of planets.  These three factors contribute to the conclusion that globular clusters are not optimal places to search for transiting planets.  
 
\section{Impact of tidal evolution on the Kepler Open Cluster Survey}
\label{sec:kepler}
Since metallicity, cluster dynamics, and tidal evolution can alter the population of hot Jupiters in 47~Tuc, we next investigate the possibility that the signature of tidal evolution can be observed in the less dense and more metal rich environs of open clusters.  In the field, the frequency of hot Jupiters as a function of stellar age points to tidal evolution \citep{jackson09}, but is complicated by highly uncertain stellar ages for most main sequence field stars.  Open clusters, however, are reasonably homogenous populations of coeval stars.  If a large sample of open clusters of varying ages were observed for transits in a consistent way, the distribution of hot Jupiters in orbits $<$0.1~AU could be used to test tidal evolution models.

The {\em Kepler} mission may provide a unique opportunity to do that.  Within the {\em Kepler} field there reside four open clusters with a wide spread in age: NGC 6791 (t=7.9~Gyr), NGC 6811 (t=0.6~Gyr), NGC 6819 (t=2.2), and NGC 6866 (t=0.4~Gyr).  If enough cluster members can be observed for transiting planets, significant differences in the planetary populations can be ascertained, either in a dearth of observed hot Jupiters or in the distribution of planets as a function of orbital period.

In anticipation of data on these clusters from {\em Kepler}, we performed our tidal evolution and survey calculations for each cluster.  We obtained the estimated cluster age, distance from the Earth, and cluster metallicity from the WEBDA database.  We then used that information to determine the brightest and dimmest main sequence cluster members where {\em Kepler} can securely detect a hot Jupiter.  We converted those magnitude limits into $M_\star$ and $R_\star$ by using Yale isochrones for solar metallicity in the case of NGC 6811, NGC 6819, and NGC 6866 and [Fe/H]=0.4 for NGC 6791 \citep{an07,an09}.  Thus we have made our calculations based on a minimum stellar mass and a maximum stellar mass in each cluster.  Again, in some clusters, the brightest stars have significantly larger radii than their ZAMS values.  To correct for this we used the youngest Yale isochrones with the same metallicity to determine primordial R$_{\rm {ZAMS}}$.  To determine R$_\star$ we used isochrones at the age of the clusters.  Using the expected {\em Kepler} relative photometry long cadence accuracies quoted in the {\em Kepler} GO Program handbook, we determine that one can reliably detect a  Jupiter mass planet for a $\sim$1$M_\odot$ star with V$<$18.3.  

Following our procedure for the general survey and the HST survey of 47~Tuc, we pick an observing window and observing cadence.  The long cadence exposures are sums of 270 integrations with an exposure time of 6.02s and a readout between sequences of 0.52s.  Thus there is a sample of a target's light-curve every thirty minutes.  We assume an observing window of 60 days, and require a $\sigma_{\rm {tran}}\ge$6 detection with at least 2 transits, which means planets with periods $<$30 days can be detected.  

\begin{figure}
\plotone{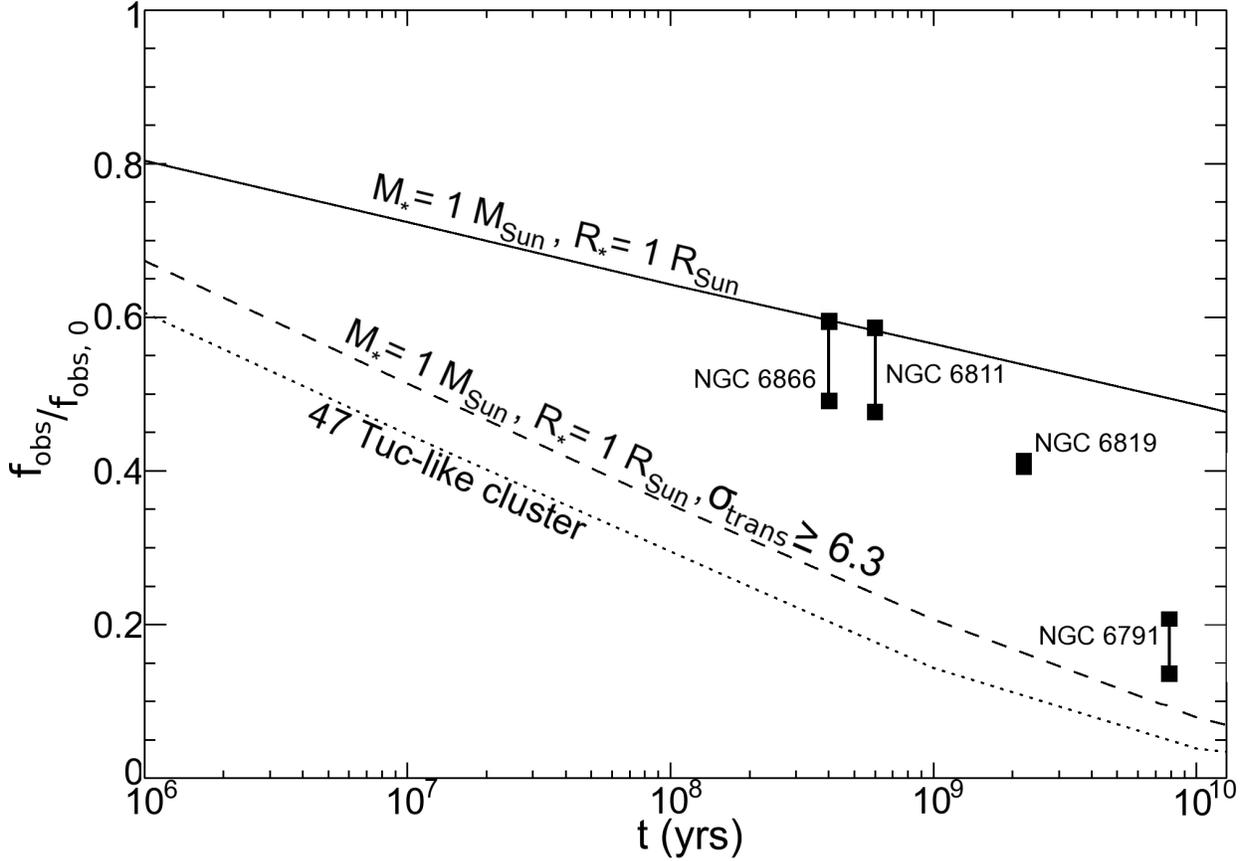}
\caption{\label{fig:f4} Expected detectable fraction of planets relative to a population that does not undergo tidal evolution $f_{obs}/f_{obs,0}$ as detailed in Sections \ref{sec:tuc} and \ref{sec:kepler}.  The solid line at the top represents an integration over semi-major axis of panel a) of Figure \ref{fig:f1}  for a sun-like cluster of stars and idealized observing conditions.  The dashed line corresponds to a cluster of Sun-like stars with more realistic observing conditions and corresponds to panel b) of Figure \ref{fig:f1}.  The dotted line corresponds to the 47~Tuc HST survey conditions and  Figure \ref{fig:f3}.  Overplotted are our estimates of $f_{obs}/f_{obs,0}$ for clusters within the {\em Kepler} FOV, using values discussed in Section \ref{sec:kepler}.}
\end{figure}

Figure \ref{fig:f4} shows the results of Monte Carlo simulations with {\em Kepler}, giving the relative fractions of remaining hot Jupiters with periods of $<$ 30 days compared to no tidal evolution ($f_{obs}/f_{obs,0}$).  For each cluster we have a range of ($f_{obs}/f_{obs,0}$) based upon the brightest and dimmest cluster members accessible with {\em Kepler} giving a range in ($f_{obs}/f_{obs,0}$) one could expect from a full survey with a distribution of stellar masses and radii.  

Given the long observing window and the high precision, most of the expected fractional yields of the clusters lie close to our solid curve in Figure \ref{fig:f4}, primarily due to the high photometric accuracy of {\em Kepler} relative to the signal of a hot Jupiter transit.  The exception is NGC~6791, primarily because of its larger distance and larger age.  A survey of this cluster is primarily sensitive to only the shortest period orbits and is thus heavily affected by tidal evolution.  Ground based surveys of NGC 6791 observed roughly 3300 cluster members for transiting planets, with an expectation of 2-3 planets observed \citep{montalto07}.  The null result of that survey would suggest that planets in that cluster could be $<$1/3 as frequent compared to field stars.  This is consistent with our calculations that when tidal evolution is taken into account, one would expect only 15\% as many planets compared to field stars.

\begin{figure}
\plotone{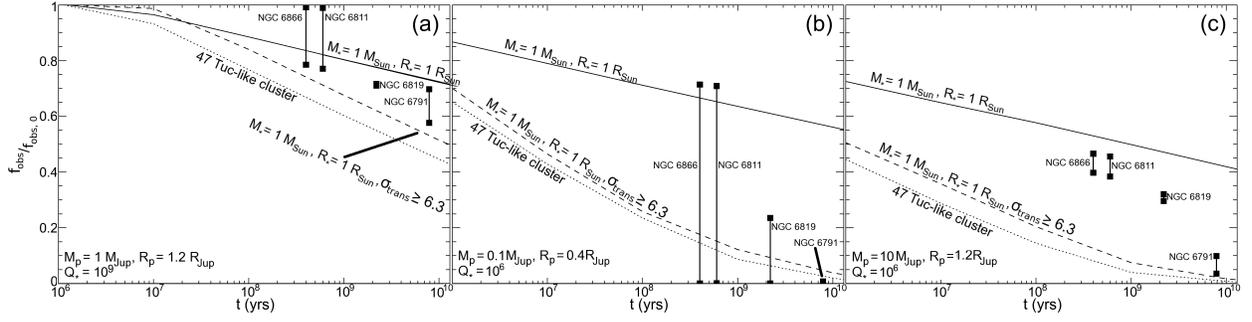}
\caption{\label{fig:f5} Same as for Figure \ref{fig:f4} but with different assumed parameters.  Panel a) shows the resulting detectable fraction of planets for a $Q_\star$=10$^9$, Panel b) shows the resulting detectable fraction for target planets with M=0.1~M$_J$, and Panel c) shows the resulting detectable fraction for target planets with M=10~M$_J$.  Only the brightest cluster members of NGC~6866, NGC~6811, and NGC~6819 will have detectable transiting M=0.1~M$_J$ planets.}
\end{figure}

Figure \ref{fig:f5} shows how these predictions might change if Q$_\star^\prime$ or M$_p$ change.  In general, the resulting $f_{obs}/f_{obs,0}$ follow the changes in Figure \ref{fig:f2}.  Increasing Q$_\star^\prime$ decreases the effect tidal evolution has on a transit survey, while lower mass planets will be difficult to observe with {\em Kepler} given the required S/N.  In fact, for the dimmer stars, Kepler will not be sensitive to smaller planets.  Higher mass planets will be about equally observable after tidal evolution. 

Figure \ref{fig:f6} shows the different expected distributions of planets as a function of semi-major axis for the four clusters at the expected ages.  The main feature of tidal evolution is an increasing minimum cut-off period as the cluster age increases as well as an increasing period of the planets most likely to be detected.   With more information on the open cluster members in the {\em Kepler} fields, we could produce a more accurate prediction for the number of planets expected to be discovered given a specific observing strategy.

\section{Discussion}
\label{sec:disc}

We have shown that the tidal evolution of hot Jupiters can have a significant impact on the observability of transiting planets in older open and globular clusters.  These impacts are 1) fewer expected hot Jupiters with increasing cluster age, and 2) an increasing minimum period for detected planets with increasing cluster age.  The loss of close-in planets has a direct effect on transit surveys that are optimized for large radius, short period planets, because these are the planets that are lost the quickest.  Transit surveys sensitive to less massive planets with longer orbital periods should therefore be more efficient at finding planets than those sensitive to only larger, close in planets.

This impact could explain the relative dearth of transiting planets in cluster surveys, even when a few detections are expected from the surveys in aggregate.  For clusters with ages $>$ 1~Gyr, one would expect at least 1/3 as many hot Jupiters as were originally formed.  Beyond $>$10~Gyr, less than a few percent of the original population remain.

Our models have important assumptions and uncertainties.  In our models, we've ignored any effect of planetary mass loss on orbital evolution. As orbital decay brings a planet closer to its host star, the planet will fill more and more of its Roche radius. For example, a planet with Jupiter's mass and a radius of 1.2 Jupiter radii in orbit around a Sun-like star will fill its Roche radius when $a \simeq 0.009$ AU, so the planet will encounter this distance before the stellar radius. Nearing this orbital distance, the planet may begin to lose mass as a result of Roche-lobe overflow. In this case, exchange of momentum between the escaping mass and the remaining planet may modify the orbital evolution from what we've assumed here, even perhaps causing the orbit to expand \citep{gu}.  This process might reduce the frequency of tidal destruction, revising our results. This process might also completely strip the atmospheres of close-in gas giants, stranding their rocky and icy cores in more distant orbits, and subsequent orbital evolution of the remnant cores would be much slower than for the original planets \citep[][in press]{jackson09b}. Some of these outcomes can be observationally tested.  If planets are engulfed, one would expect a signature of pollution in the stellar atmosphere \citep{sandquist02}, or an increase in stellar rotation rates \citep{massarotti08}.  If the process strips just the envelope but leaves a dense core, there might be an excess of $\sim$5-10 M$_{E}$ planets with short periods above that expected through orbital migration alone.  There might even be a mass-period relationship for the remnant cores, analogous to that observed for tidally-stripped white dwarfs \citep[e.g.][]{rappaport95}. Although transit surveys may have difficulty detecting these small stranded remnant cores, their detection would provide an important clue to the fate of close-in gaseous planets.

Our results depend on our choice of $Q_\star^\prime$.  In addition to the exact value of $Q_\star^\prime$ being uncertain, $Q_\star^\prime$ may change as a function of stellar type \citep{barker10}. The value of $Q_\star^{\prime}$ may depend in a complex way on the tidal frequencies in a physical situation, and remains poorly understood.  In any case, our results for a given $t$-value can be adjusted for different $Q_\star^{\prime}$-values by multiplying our chosen $t$ by $Q_\star^{\prime}/10^6$.

We've assumed that the orbital inclination relative to the stellar rotation axis is effectively zero. While tides raised on a host star can reduce any misalignment between the planet's orbital plane and stellar equator, this reduction would still result in a random alignment of orbital planes along an observer's line-of-sight. If the inclination of a star's equator relative to the observer's line-of-sight is determined to be small, the probability that an orbiting planet may be observed to transit may be much larger \citep{beatty10}. In those cases, tidal decay of planetary orbits might increase a planet's transit probability over time.  

Recent Rossiter-McLaughlin (R-M) results show that a growing number of transiting planets have significant misalignments between the orbital axis and the stellar rotation axis as projected onto the sky, including some fully retrograde orbits \citep{hebrard08,pont09,winn09,anderson10,jenkins10}.  Recent statistical analyses of R-M measurements suggest that there are two distinct populations of hot Jupiters, $\sim$2/3 of hot Jupiters are part of a population with a mean value similar to the Sun's obliquity with the Solar System, and $\sim$1/3 of hot Jupiters come from an isotropic distribution \citep{fabrycy09}.  Our tidal evolution model does not include spin-orbit misalignments in the timescales for destruction, but \citet{barker09} show that misalignments tend to speed up tidal evolution, by as much as a factor of $\sim$2 for retrograde orbits.  However, given the rapid destruction of hot jupiters at small orbital separations anyway, a factor of two increase in the rate should not significantly impact our results.

We've also assumed orbital eccentricity of our model planets are small enough to be negligible. If the orbital eccentricity were larger, then the tide raised on the planets by the host stars would increase orbital decay rates. Without additional perturbations, tides can damp orbital eccentricity more quickly than orbits decay \citep{rasio96,jackson09,barker09}, so even if eccentricities start out non-negligible, they should become small relatively quickly. However, if additional companions (either stars or planets) are present, gravitational perturbations from those companions may keep orbital eccentricity non-negligible for much longer than in the absence of such perturbations \citep[e.g.][]{mardling07}, enhancing orbital decay rates.

Despite these uncertainties, tides play a key role in the evolution of hot Jupiters and their effect has the potential to be directly observed.  In addition, since transit surveys are affected by tidal evolution, observers wishing to discover transiting planets in clusters must have sufficient accuracy to detect lower mass planets, larger samples, or longer observation windows to be confident that a significant number of detections will occur.  

\acknowledgements We would like to thank Ron Gilliland, Steinn Sigurdsson, Rory Barnes, and the anonymous referee for helpful comments and suggestions on the manuscript.  This research was supported by an appointment to the NASA Postdoctoral Program at the Goddard Space Flight Center, administered by Oak Ridge Associated Universities through a contract with NASA.  This work made use of the Extrasolar Planets Encylopedia at exoplanets.eu.

\bibliography{scibib}
\bibliographystyle{apj}

\end{document}